\documentstyle[12pt,epsf]{article}
\def\QED{\mbox{\rule[0pt]{1.5ex}{1.5ex}}}

\def\appendix{\par
    \setcounter{section}{0}\setcounter{subsection}{0}
    \def\thesection{\Alph{section}} \section*{Appendix}
}

\newenvironment{SUMMARY}{\noindent\small\rm\relax
{\small\bf\relax SUMMARY}\hskip1pc}{\par}
\newenvironment{keywords}{\noindent\small\it\relax{\small\it\relax key words:}}{\par}

\newcommand{\tr}{\mbox{\rm Tr\,}}
\renewcommand{\epsilon}{\varepsilon}
\renewcommand{\phi}{\varphi}
\newcommand{\Ent}{{\rm Ent}}
\newcommand{\I}{{\rm I}}
\newcommand{\D}{{\rm D}}

\def\argmax{\mathop{\rm argmax}}

\begin{document}

\title{Numerical Experiments on The Capacity of Quantum Channel with Entangled Input States}
\author{Susumu Osawa \\
{\small High Energy Accelerator Research
Organization (KEK),  }\\
{\small Tsukuba, Ibaraki 305-0801, Japan.}\\
{ \small E-mail: osawa@post.kek.jp}\\
{\small and}\\
Hiroshi Nagaoka\\
{\small  Graduate School of Information Systems,}\\
{\small University of Electro-Communications,}\\
{\small Chofu, Tokyo 182--8585, Japan.}\\
{ \small E-mail:  nagaoka@is.uec.ac.jp}
}

\date{}

\maketitle
\begin{SUMMARY}
The capacity of quantum channel with product input states was formulated
by  the quantum coding theorem. However,
whether entangled input states can enhance the quantum channel is still
open.
It turns out that this problem is reduced to a
special case of the more general problem whether the capacity of product
quantum channel
exhibits additivity. In the present study,
we apply one of the quantum Arimoto-Blahut type algorithms to the latter
problem.  The results suggest that the additivity of product quantum
channel
capacity always
holds and that entangled input states cannot enhance the quantum  channel
capacity
\footnote{The content of this paper was partly presented at 
the second QIT \cite{nagaoka2} and the 
22nd SITA \cite{osawa}.}.
\end{SUMMARY}


\section{Introduction}\label{intro}
 The coding theorem for quantum channels\footnote{
In this paper we treat only quantum memoryless channels 
and simply call them quantum channels.}
 was proved in  recent publications 
 \cite{Holevo2,Schumacher1} combined with pioneering works
 such as \cite{Holevo0,Holevo1}.
This theorem 
gives the formula of the capacity of quantum channel with product
(not
entangled)
input states.
 On the other hand, the use of entangled states in quantum communications
provides us with another interesting aspect,
which was already pointed out in \cite{Bennett}.
Even though lots of theoretical  attempts have been made so far in this
direction
(see section \ref{sec:related_works} for recent works 
on this subject),
we are still far from deep understanding of
the role of entanglement in quantum communications.
In particular,
whether the use of entangled input states can increase the capacity of
quantum 
channel is a big open problem, which, as is seen from Theorem 1 below,
can be reduced to a special case of another open problem
whether the capacity of product quantum channel exhibits additivity.
In the present study, we examine the latter problem numerically
by means of a quantum version of Arimoto-Blahut algorithm \cite{nagaoka1},
and
observe that the additivity seems to always hold. 

\section{Capacity of quantum channel and entangled states}

\subsection{Quantum channel with product input states}

In this section,
we give a brief review of 
the standard notion of quantum channel with product input states
and its capacity. 

Let ${\cal H}$ be a Hilbert space which corresponds to a quantum system.  A
quantum
state is represented by a density operator on  ${\cal H}$, i.e. non-negative
operator
with unit trace.
We denote by ${\cal  S}({\cal H})$ the totality of density operators on
${\cal H}$.
Letting ${\cal H}_1$ and ${\cal H}_2$ be input and output systems, a quantum
channel
is described by a
completely positive \cite{Stinespring} trace preserving linear map 
\[\Gamma \; : \;{\cal T}({\cal H}_{1})\to{\cal T}({\cal H}_{2})\]
where ${\cal T} ({\cal H}_1)$ and   ${\cal T} ({\cal H}_2)$ are the
totalities of 
the trace class operators on ${\cal H}_1$ and ${\cal H}_2$. 
Note that the complete positivity and the
trace preservation  jointly characterize the physical realizability of quantum channels
\cite{Kraus}.

 A quantum communication system in which  a quantum channel  $\Gamma$ 
is used
$n$ times is described as follows.
A message set ${\cal M}_n :=\{1,2, \cdots , M_{n}\}$ denotes the totality of
the
messages which are to be
transmitted.  Each message  $k\in{\cal M}_n$ is encoded to a codeword 
which is a product state in the form
$\rho^{(n)}(k):=\rho_{1}(k) \otimes \cdots \otimes  \rho_{n}(k)$ on ${\cal
H}_{1}
^{\otimes n},$ where ${\cal H}_{1}
^{\otimes n}$ denotes the tensor product Hilbert space ${\cal H}_{1}\otimes
\cdots
\otimes {\cal H}_{1}$. The
sender transmits the codeword by multiple use of a quantum channel $\Gamma$.
Then
the received state is also a
product state $\Gamma^{\otimes n}(\rho^{(n)}(k))=  \Gamma (\rho_{1}(k))
\otimes
\cdots  \otimes
\Gamma(\rho_{n}(k))$ on ${\cal H}_{2} ^{\otimes n}$. Here $\Gamma ^{\otimes
n}$ denotes the n-fold tensor
product channel $\Gamma \otimes \cdots \otimes \Gamma$ acting on ${\cal
T}({\cal
H}_{1} ^{\otimes n})$.
 The receiver estimates which codeword has been actually transmitted by
performing
an ${\cal M}_n$-valued
measurement.  Mathematically, this measurement is represented by a positive
operator
valued measure (POVM)
$X^{(n)}=\{X_1^{(n)}, \cdots  ,X_{M_n}^{(n)} \}$ on ${\cal H}_{2} ^{\otimes
n}$, i.e.
$X_k^{(n)} \ge 0 \,
(k=1, \cdots ,M_{n})$ and $\sum_{k=1}^{M_n}X_k^{(n)}=I$, where $I$ denotes
the
identity operator on ${\cal H}_{2} ^{\otimes n}$.

Given a {\it coding system} ${\Phi_n}$ consisting of
codewords $\{\rho ^{(n)}(k)\}_{k=1}^{M_n}$ and a measurement  $X^{(n)}$,
the error
probability averaged over all codewords
is given by
\begin{equation}
\label{Per}
    P_{er}(\Phi_n,\Gamma)=1-\frac{1}{M_n}\sum_{k=1}^{M_n}\tr[\Gamma^{\otimes
n}( \rho ^{(n)}(k)) X_k^{(n)}],
\end{equation}
and the quantity $R_n(\Phi_n):= \log M_n /n$ is  called the {\it rate} of
the coding
system $\Phi_{n}$.
Now the capacity of the quantum channel $\Gamma$ with product input states
is
defined as
\begin{equation}
  C(\Gamma ) := \sup_{\{\Phi_n\}}
 \{  \lim_{n \to \infty  } R_n(\Phi_n) \; ; \,
      \lim_{n \to \infty } P_{er}(\Phi_n, \Gamma)=0 \}.    \label{capdef}
\end{equation}

 Next let us introduce the quantum mutual information. Let
 \begin{eqnarray*}\Pi_{n} &:=& \{ (\lambda_{1},\cdots ,\lambda_{n}\, ;\,
\sigma_{1},\cdots ,\sigma_{n})\;;  \\
&&
\quad  0 \le \lambda_i \in {\bf R}, \; \sum_{i=1}^{n} \lambda_{i} =1,\;
\sigma_i \in
{\cal S}({\cal H}_{1})\},
\\
 \Pi &:=& \bigcup_{n}^{\infty} \Pi_{n}.
\end{eqnarray*}
An element $\pi =(\lambda_{1},\cdots ,\lambda_{n}\, ;\,
\sigma_{1},\cdots ,\sigma_{n})\in\Pi$
is considered as a
discrete probability distribution  on ${\cal S}({\cal H}_1)$ assigning
probability
$\lambda_i$ to the state
$\sigma_i$  for each $i$.  The
quantum mutual information for
$\pi$ and
$\Gamma$ is then defined as
\begin{equation}
 \I ( \pi ; \Gamma):= \sum_{j}\lambda_{j} \D (\Gamma (\sigma_j)
\|\Gamma(\rho)),
\end{equation}
where $\rho :=\sum_{j}\lambda_{j}\sigma_j$ is a convex  combination of the
states
in $\pi$ and $\D ( \rho \|\sigma ):=\tr [\rho (\log \rho - \log \sigma )]$
is the
quantum relative entropy.

Now, the quantum channel coding theorem
\cite{Holevo0,Holevo1,Holevo2,Schumacher1}
states that
\begin{equation}
C(\Gamma) = \sup_{\pi \in \Pi} \I( \pi ;  \Gamma).  \label{qcapacity}
\end{equation}
In addition, supremization is reduced to maximization on certain
finite-dimensional
compact
set \cite{fujiwara}.  That is,
\[C(\Gamma)=\max_{\pi \in \Pi_{n}} \I(\pi ; \Gamma)=\max_{\pi \in
\Pi_{n}^{e}}\I(\pi ; \Gamma) \]
where $n = \dim \Gamma ({\cal S}({\cal H}_1))+1$,
$ \Pi_{n}^{e}:=\{ (\lambda_i ;\sigma_i)\in \Pi_n \;;\; \sigma_i \in \partial
_{e}{\cal S}({\cal H}_1),\, i=1, \cdots ,n \}$.
Here $\partial _{e}{\cal S}({\cal H}_1)$ is the  totality of extreme points
(pure
states) of ${\cal S}({\cal H}_1)$.

\subsection{Quantum channel with entangled input states}

Some states on a tensor product Hilbert space cannot be represented as
product states
or
their convex combinations. These states are called {\it entangled states}.
In the formulation given in the previous section we treated only
product states as inputs to (and, consequently, outputs from) a quantum
channel.
Now let us consider communication
systems in which we are allowed to use entangled input states.

The capacity  of the quantum channel $\Gamma$
with entangled input states $\tilde{ C}(\Gamma)$
is defined in the same way as  (\ref{capdef}) except that arbitrary states
on ${\cal
H}_1^{\otimes n}$, not necessarily product states,
can be codewords.
It is obvious by
definition that
\begin{equation}
\tilde{C}(\Gamma) \ge C(\Gamma).  \label{entgenoent}
\end{equation}
However, neither example of
channel exhibiting the strict inequality
nor proof that the equality
\begin{equation}
\tilde{C}(\Gamma) = C(\Gamma)   \label{enteqalnoent}
\end{equation}
always holds have been reported yet\footnote{
An earlier example of statement of this problem
is found in the concluding remarks of \cite{fujiwara}.
}.
This problem can be reduced to the additivity
problem for the capacity of
product channels as described in the next section.

\subsection{Capacity of product quantum channel}
\label{product_channel}
Let $\Gamma^{(i)} \, : \,{\cal T}({\cal H}_{1}^{(i)})\to{\cal T}({\cal
H}_{2}^{(i)})$
for $i=1, 2$ be quantum
channels, and let $\Gamma^{(1)} \otimes \Gamma^{(2)} \, : \,{\cal T}({\cal
H}_{1}^{(1)} \otimes {\cal
H}_1^{(2)})\to{\cal T}({\cal H}_{2}^{(1)}\otimes {\cal H}_{2}^{(2)})$ be
their
product channel. The capacity
$C(\Gamma^{(1)} \otimes \Gamma^{(2)})$  is defined as (\ref{capdef}) by
replacing
$\Gamma$ with $ \Gamma
^{(1)}\otimes \Gamma^{(2)}$ in which each input state
(code word) is written in the form $\rho^{(n)}(k)=\rho_{1}(k) \otimes
\cdots \otimes \rho_{n}(k)$, where $\rho_{i}(k)\;(i=1,\cdots ,n)$ are
arbitrary
states  on ${\cal H}_1^{(1)}
\otimes {\cal H}_1^{(2)}$. Then it is easy to see that the superadditivity
\begin{equation}
 C(\Gamma^{(1)} \otimes \Gamma^{(2)})\ge C(\Gamma^{(1)}) + C(\Gamma^{(2)})
\label{superadditive}
\end{equation}
holds. However, as in the case of (\ref{entgenoent}), we have no example
of the strict inequality nor proof that the additivity
\begin{equation}
 C(\Gamma^{(1)} \otimes \Gamma^{(2)})= C(\Gamma^{(1)}) + C(\Gamma^{(2)})
\label{additive}
\end{equation}
always holds.   
Actually, this problem includes the 
previous one as is seen from the following theorem,
whose proof is  given in Appendix A.

\bigskip\par\noindent
{\bf Theorem 1:}
\begin{equation}
\tilde{C}(\Gamma)=\lim_{N \to \infty} \frac{C(\Gamma ^{\otimes
N})}{N}=\sup_{N}\frac{C(\Gamma ^{\otimes N})}{N}
\end{equation}
\noindent
holds.
Here $C(\Gamma ^{\otimes N})$ is defined as (\ref{capdef}) by replacing
$\Gamma$ with
$ \Gamma ^{\otimes N}$ in which  each input state is written in the form
$\rho^{(n)}(k)=\rho_{1}(k) \otimes \cdots \otimes \rho_{n}(k)$, where
$\rho_{i}(k)\;(i=1,\cdots ,n)$ are arbitrary states  on ${\cal H}_1^{\otimes
N}$.
\bigskip\par\noindent
If the additivity (\ref{additive}) always holds, we have 
$C(\Gamma ^{\otimes N}) = N C(\Gamma)$, which, combined with Theorem 1,
leads to the equality (\ref{enteqalnoent}). In other words, the additivity
implies that
entanglement of input states cannot increase
the capacity of quantum channel.

\subsection{The aim of the present paper and related works}
\label{sec:related_works}

In the last few years  the additivity (\ref{additive}) 
has gradually been receiving recognition 
as a difficult but important 
problem in the quantum information theory, 
and has been proved for several special cases. 
At the moment, proofs are known for the cases when 
$\Gamma^{(1)}$ is arbitrary and $\Gamma^{(2)}$ is the identity
\cite{Schumacher2}, when  $\Gamma^{(1)}$ is arbitrary and 
$\Gamma^{(2)}$ is  either a Holevo's classical-quantum
or quantum-classical channel \cite{King3} and 
when $\Gamma^{(1)}$ is arbitrary and 
$\Gamma^{(2)}$ is a certain class of
unital binary channels \cite{King4}.  
See also \cite{Bruss,Holevo3}, whose results  are now 
included in some of the above-mentioned ones.  
On the other hand, no example of 
channel violating the additivity have been found so far, 
and naturally 
the conjecture that the additivity always holds is arising 
\cite{Amosov1,Holevo3,King2,King3,nagaoka2,osawa,Schumacher2}. 
The aim of the present paper is to show that 
an efficient algorithm for computing quantum channel capacity, 
which was recently introduced by 
one of the authors, is applicable to verification of the conjecture 
and  to report that all the randomly chosen channels have exhibited the additivity
\footnote{
In several references such as 
\cite{Amosov1,King1,King2} 
it is shortly mentioned, without any detail, that 
some numerical works relating the conjecture 
have been made.}.

\section{Numerical experiments on the additivity}
\subsection{Quantum version of Arimoto-Blahut algorithm}
\label{sec:Arimoto-Blahut}
The Arimoto-Blahut algorithm is known for computing the capacity of
classical channel
\cite{Arimoto,Blahut}.
Recently, one of the authors proposed some algorithms of this type for
computing the
capacity of quantum channel
\cite{nagaoka1,nagaoka2}.  We use one of these, which is called the {\it
boundary
algorithm}
since its
recursion works on the extreme boundary set
$\partial _{e}{\cal S}({\cal H}_{1})$. The outline of the theoretical basis
is as
follows.

Let us introduce a two-variable extension of $\I(\pi;\Gamma)$:
 \begin{equation}
 J(\pi, \pi^{\prime}) := -\D(\lambda \| \lambda ^{\prime})+\sum_{i=1}^{n}
\lambda_{i}\tr [ \Gamma (\sigma_{i})\Phi
(\sigma_{i}^{\prime},\rho^{\prime})],
\end{equation}
where
\[ \pi =(\lambda_i; \sigma_i),\;\pi^{\prime} =(\lambda_i^{\prime};
\sigma_{i}^{\prime}) \in \Pi_n,  \]
\[ \D(\lambda \| \lambda ^{\prime}):= \sum_{i=1}^{n}\lambda_i \log
\frac{\lambda_i}{\lambda_i^{\prime}}, \;
\rho^{\prime}:=\sum_{i=1}^{n}\lambda_i^{\prime}\sigma_i^{\prime}, \]
\[ \Phi (\sigma_{i}^{\prime},\rho^{\prime}):= \log (\Gamma
(\sigma_{i}^{\prime}))-\log (\Gamma (\rho^{\prime})). \]
Then it holds that
\begin{equation}
\I( \pi ; \Gamma) = J(\pi , \pi)= \max_{\pi^{\prime}}J(\pi , \pi^{\prime}).
\label{j1}
\end{equation}
In addition, we can compute $\displaystyle\hat{\pi}=(\hat{\lambda_i}$,
$\hat{ \sigma_i}) := \argmax_{\pi}
J(\pi,\pi^{\prime}) $ by the following equations.
\[
\hat{ \sigma_i}= \argmax_{\sigma \in {\cal S}({\cal H}_{1})} \tr [ \Gamma
(\sigma)\Phi (\sigma_{i}^{\prime},\rho^{\prime})],
\]
\[
\textstyle
\hat{\lambda_i}=\lambda_i^{\prime} \exp ( \tr [ \Gamma
(\hat{\sigma_{i}})\Phi
(\sigma_{i}^{\prime},\rho^{\prime}) ])/\hat{Z},
\]
where $\hat{Z}$ is the normalizing constant:
\[
\hat{Z}:= \sum_{i=1}^{n}\lambda_i^{\prime} \exp ( \tr [ \Gamma
(\hat{\sigma_{i}})\Phi (\sigma_{i}^{\prime},\rho^{\prime}) ] ).
\]
Note that, since $ \tr [ \Gamma (\sigma)\Phi
(\sigma_{i}^{\prime},\rho^{\prime})]$ is
linear in $\sigma$, we can always choose $\hat{\sigma_i}$
to be  an extreme point of ${\cal S}({\cal H}_1)$, i.e. a pure state $|
\psi_i\rangle
\langle \psi_i|$,  where
$|\psi_i\rangle$ is a normalized eigenvector of $\Gamma^{\ast}(\Phi
(\sigma_{i}^{\prime},\rho^{\prime}))$
corresponding to the maximum eigenvalue.
Here $\Gamma^{\ast}\;:\;{\cal B}({\cal H}_2) \to {\cal B}({\cal H}_1)$
denotes
the dual map of $\Gamma$ defined by
$\tr[\Gamma(X)Y]=\tr[X\Gamma^{\ast}(Y)]$ for $\forall X \in
{\cal T}({\cal H}_1)$ and $\forall Y \in {\cal B}({\cal H}_2)$, where ${\cal
B}({\cal H}_i)\;(i=1,2)$ are the
totalities of bounded operators on ${\cal H}_i$.

Given a number $n\le\dim \Gamma ({\cal S}({\cal H}_1))+1$
 and an arbitrary  initial element $ \pi^{(1)} \in \Pi_{n}$,
let the sequence
$\{\pi^{(k)}\}_{k=1}^{\infty}$ be defined by
\begin{equation}
 \pi^{(k+1)}:= \argmax_{\pi}J(\pi, \pi^{(k)}).
\end{equation}
Note that the sequence $\{\I(\pi ^{(k)}; \Gamma)\}_{k=1}^{\infty}$ is
monotonous,
since
\begin{equation}
\I(\pi ^{(k)}; \Gamma) \le J(\pi ^{(k+1)}, \pi ^{(k)}) \le \I(\pi ^{(k+1)};
\Gamma)
\end{equation}
holds.  Therefore we can efficiently compute the limit value $\lim_{k \to
\infty}\I(\pi ^{(k)}; \Gamma)$.  Unfortunately, it is not necessarily the
quantum
channel capacity since the  quantum version of Arimoto-Blahut algorithm does
not
assure the global maximum. Thus we make several convergent sequences and
adopt the
maximum limit value as an estimate of the capacity. We judge that a sequence
reaches 
the limit value when ten successive numerical values are the same to six
places of
decimals.

\subsection{Setting of the experiments}

\begin{table*}[t]
\caption[character1]{Examples of quantum binary channels $(A,b)$}

\begin{center}
\tabcolsep= 3pt \footnotesize
    \begin{tabular}{|c|c|c|} \hline
               &  A & b \\ \hline
  $\Gamma_1$   & $ \pmatrix{ 0.5 & 0 & 0 \cr
                             0 &  0.4 & 0 \cr
                             0 & 0 &  0.2 \cr
                          }$
&    $\pmatrix{0.2 \cr
                 0  \cr
                 0 \cr
                } $     \\ \hline

  $\Gamma_2$   & $ \pmatrix{ 0.05 & -0.2 & 0.4 \cr
                             -0.2 &  -0.05 & -0.2 \cr
                             0.2 & 0 &  -0.5 \cr
                          }$
&    $\pmatrix{0 \cr
                 0  \cr
                 0.1 \cr
                } $     \\ \hline

  $\Gamma_3$   & $ \pmatrix{ 1/\sqrt{2}   &-1/\sqrt{6}  &1/\sqrt{3}  \cr
    1/\sqrt{2}       & 1/\sqrt{6}   & -1/\sqrt{3}\cr
      0  & 2/\sqrt{6}   &1/\sqrt{3} \cr
                          }\cdot
\pmatrix{  -0.45 & 0 & 0 \cr
             0 &  0.6 & 0 \cr
             0 & 0 &  -0.6 \cr
                          }\cdot
\pmatrix{ 0.8   &  0.6 & 0 \cr
           0.6 &  -0.8 & 0 \cr
           0 & 0 &  1 \cr
}$
&    $\pmatrix{0.2 \cr
                 -0.2  \cr
                 0.2 \cr
                } $     \\ \hline
  $\Gamma_4$   & $ \pmatrix{ 0.1 & -0.3 & 0 \cr
                             -0.3 &  -0.1 & -0.2 \cr
                             0 & 0 &  -0.05 \cr
                          }$
&    $\pmatrix{0 \cr
               0.2  \cr
                 0.55 \cr
                } $     \\ \hline
            \end{tabular}
          \end{center}

\label{chara1}
\end{table*}%

\begin{table*}
\caption[character2]{Examples of quantum channels with ${\cal H}_1={\cal H}_2={\bf C}^3$ having generators
of the form $\{V_1, V_2, \sqrt{I-V_1^{\ast}V_1-V_2^{\ast}V_2} \}$ }

   \begin{center}
\tabcolsep= 3pt \footnotesize
    \begin{tabular}{|c|c|c|} \hline
             &  $V_1$ & $V_2$ \\ \hline
  $\Gamma_5$   & $ \pmatrix{ 0.2 & 0.3 & 0.4 \cr
                             0 &  0.5i & 0 \cr
                             0.1i & 0.4i &  0.5i \cr
                          }$
&    $\pmatrix{0.1-0.3i & 0 & 0 \cr
                 0      &  -0.3i & 0.1-0.2i \cr
                 0.3-0.3i & 0.2+0.1i & 0\cr
                } $     \\ \hline

  $\Gamma_6$   & $ \pmatrix{ 0.19 & 0.7 & -0.1+0.3i \cr
                             0.4i &  0.06 & -0.1+0.05i \cr
                             0.2 & 0.39 &  0.4-0.4i \cr
                          }$
&    $\pmatrix{0.3 & -0.1 & 0.1\cr
               0.2 & 0.3 & 0.02i  \cr
               0.1 & 0.2 & 0.1i   \cr
                } $     \\ \hline
  \end{tabular}
  \end{center}

\label{chara2}
\end{table*}%

 We apply the quantum Arimoto-Blahut algorithm 
to various quantum channels $\Gamma^{(1)}$ and 
$\Gamma^{(2)}$ 
as well as their product channels 
$\Gamma^{(1)}\otimes\Gamma^{(2)}$ 
to examine whether the additivity (\ref{additive}) 
always holds. 
Here we restrict ourselves to the case when  
${\cal H}_1 ={\cal H}_2 ={\bf C}^2$ or 
${\cal H}_1 ={\cal H}_2 ={\bf C}^3$ 
to reduce the computational complexity. 
Tables \ref{chara1} and \ref{chara2}   
show representative 
examples of channels used as $\Gamma^{(1)}$ 
and  $\Gamma^{(2)}$ in the experiments. 
   Here the first four channels $\Gamma_1, \cdots, \Gamma_4$ 
are quantum binary channels in the sense that 
${\cal H}_1 ={\cal H}_2 ={\bf C}^2$, and 
are expressed in terms of the coefficients
$(A,b)$ 
of the corresponding
affine transformations on ${\bf R}^3$ (see Appendix C),
while
the other channels $\Gamma_5$ and $\Gamma_6$ 
are of ${\cal H}_1 ={\cal H}_2 ={\bf C}^3$ and 
are expressed by the generators $\{V_1,
\cdots, V_m\}$
of their operator-sum representations (see Appendix B), restricting
ourselves to
the case when $m=3$ and $V_3=
\sqrt{I-V_1^{\ast}V_1-V_2^{\ast}V_2}$.  
Note that these channels do not belong to 
the special classes mentioned in section \ref{sec:related_works} 
for which the additivity has been proved. 
These examples are chosen basically in   
random manners so that they are as generic as possible, 
not being intended to have any special properties or to
represent any concrete physical processes, 
except that some extra points are 
taken into consideration in view of 
computational efficiency and generality, as explained below. 

In the case of quantum channels with 
${\cal H}_1 ={\cal H}_2 ={\bf C}^2$, 
the necessary and sufficient 
condition for the coefficients
$(A,b)$ to represent a pseudoclassical channel is known \cite{fujiwara}. 
Here a quantum channel is said to be {\em pseudoclassical} when  
its capacity is unchanged even if the measurements $X^{(n)}$ 
on ${\cal H}_2^{\otimes n}$ 
in equation (\ref{Per}) are restricted to separable measurements 
which are constructed from the
tensor products of  measurements on ${\cal H}_2$.  Considering the 
fundamental importance of the pseudoclassicality 
in classification of quantum channels, 
we choose 
$\Gamma_1$ and 
$\Gamma_3$ to be pseudoclassical, while $\Gamma_2$ and $\Gamma_4$ to 
be non-pseudoclassical, and examine various combinations of these channels.  

In the case of quantum channels with ${\cal H}_1 ={\cal H}_2 ={\bf C}^3$,  
on the other hand, 
we do not care about the pseudoclassicality, since no 
practical criterion for this property  
is known.  
The general operator-sum representation of channel 
in this case  
is given by generators $\{V_1, \cdots, V_m\}$ satisfying 
$\sum_{k=1}^m V_k^* V_k =I$ with $m\leq 9$, while  
our setting of $\Gamma_5$ and $\Gamma_6$ 
is much more restrictive.  This restriction 
simply comes from a demand to reduce computational 
complexity.  Nevertheless, the choice of channels 
may be considered sufficiently generic in the 
sense that it does not assume any special 
structure in view of the additivity.  

As we mentioned in section \ref{sec:Arimoto-Blahut}, 
the quantum version of 
Arimoto-Blahut algorithm does not
assure the global maximum, and the limit value of $\I(\pi^{(k)} ;
\Gamma^{(1)})$ or $\I(\pi^{(k)} ; \Gamma^{(1)} \otimes \Gamma^{(2)})$
may depend on  the initial element $\pi^{(1)}\in \Pi_n$.  
Therefore,  we repeatedly apply the algorithm to a channel 
with several different initial elements, and adopt the 
maximum of the convergent values as the estimate of the capacity. 
However, it empirically appears that the 
algorithm is not so sensitive to the initial condition. 
Indeed, we have observed that randomly chosen initial conditions 
mostly yield the same convergent value as 
far as the number $n$ of the states in $\pi^{(1)}$ 
is chosen to be sufficiently large 
(i.e. $n \approx  \dim \Gamma ({\cal S}({\cal H}_1))+1$). 
The following is an example of $\pi^{(1)}$ for which 
the convergent value has attained the 
capacity when applied to each of the quantum binary channels 
$\Gamma_1, \cdots, \Gamma_4$: 
\[ \pi^{(1)}=(\lambda_1, \cdots, \lambda_4 \, ;\, \sigma_1 ,\cdots ,
\sigma_4)\]
 with $\lambda_1 = \cdots = \lambda_4 = 1/4$ and 
\[
           \sigma_1 = \frac{1}{2}\pmatrix{ 1 & 1  \cr
                     1 & 1 \cr }, \quad 
\sigma_2 = 
              \frac{1}{2}\pmatrix{ 1 & i  \cr -i  & 1 \cr }, 
\]
\[
\sigma_3 = 
                         \pmatrix{ 1 & 0  \cr
                                   0 & 0 \cr  }, \quad 
 \sigma_4 = 
   \frac{1}{2 \sqrt{3}} \pmatrix{ \sqrt{3}-1 &
-1-i    \cr
-1+i &  \sqrt{3}+1 \cr} .
\]

\subsection{Results}

\begin{table*}
\caption[result]{The capacity of the quantum channels shown in Tables 
\ref{chara1} and
\ref{chara2}  and their product channels}
\begin{center}
\tabcolsep= 3pt \footnotesize

    \begin{tabular}{|c|c|c|c|c|c|} \hline
$\Gamma^{(1)}$ & $\Gamma^{(2)}$ & $C(\Gamma^{(1)})$ & $C(\Gamma^{(2)})$ &
$C(\Gamma^{(1)})+C(\Gamma^{(2)})$  &
$C(\Gamma ^{(1)} \otimes \Gamma^ {(2)})$    \\ \hline
$\Gamma_1$ & $\Gamma_1$ &  0.138166 & 0.138166  & 0.276311 & 0.276311 \\
\hline
$\Gamma_2$ & $\Gamma_2$ &  0.258679 & 0.258679  & 0.517358 & 0.517358 \\
\hline
$\Gamma_1$ & $\Gamma_3$ &  0.138166 & 0.243068  & 0.381233 & 0.381233 \\
\hline
$\Gamma_3$ & $\Gamma_2$ &  0.243068 & 0.258679  & 0.501747 & 0.501746 \\
\hline
$\Gamma_2$ & $\Gamma_4$ &  0.258679 & 0.0898225 & 0.348501 & 0.348501 \\
\hline
$\Gamma_5$ & $\Gamma_5$ &  0.677358 & 0.677358  & 1.354716 & 1.354716 \\
\hline
$\Gamma_6$ & $\Gamma_6$ &  0.829580 & 0.829580  & 1.659160 & 1.659160 \\
\hline
$\Gamma_5$ & $\Gamma_6$ &  0.677358 & 0.829580  & 1.506938 & 1.506938 \\
\hline
\end{tabular}
\end{center}
\label{results}
\end{table*}%

\begin{table*}
\caption[limit]{The probability distributions which maximize the quantum
mutual information of the quantum channels shown in 
Tables \ref{chara1} and \ref{chara2}}
\label{piast}
\begin{center}
\tabcolsep= 3pt \footnotesize
   \begin{tabular}{|c|c|} \hline

   & $ \pi^{\ast}=(\lambda_i^{\ast}\; ; \;
\sigma_i^{\ast} )$ \\ \hline
 $\Gamma_1$ & ($0.521046, 0.478954 \;;\;$ $\pmatrix{0.500 & 0.500\cr
                                          0.500 &0.500 \cr}$,
                                          $\pmatrix{0.500 & -0.500\cr
                                          -0.500 &0.500 \cr})$ \\ \hline
$\Gamma_2$  &  ($0.512423, 0.487577 \;;\;$
                       $ \pmatrix{0.009 & 0.055-0.079i\cr
                         0.055+0.079i &0.991 \cr}$,
                       $ \pmatrix{0.991, & -0.049+0.082i\cr
                                   -0.049-0.082i &0.009 \cr}$) \\
\hline
$\Gamma_3$  &   ($0.271288, 0.728711 \;;\;$  $\pmatrix{0.00 & 0.00\cr
                                              0.00 &1.00 \cr}$,
                                             $\pmatrix{1.00 & 0.00\cr
                                                       0.00 &0.00 \cr}$) \\
\hline
$\Gamma_4$  &   ($0.47431, 0.52569 \;;\;$  $\pmatrix{0.772 & 0.398-0.133i\cr
                                                    0.398+0.133i &0.228
\cr}$,
                                  $\pmatrix{0.218 & -0.392+0.131i\cr
                                  -0.392-0.131 &0.782 \cr}$) \\ \hline
$\Gamma_5$  &   ($0.31721, 0.383025, 0.299764 \;;\;$
                           $\pmatrix{0.690& -0.365-0.260i & 0.111-0.034i \cr
                                      -0.365+0.260i & 0.290 & -0.046+0.060i
\cr
0.111+0.034i &-0.046-0.060i&0.019 \cr}$, \\
               &           $\pmatrix{0.023 & 0.079+0.024i&-0.121-0.035i
\cr
                             0.079-0.024i &0.294&-0.448+0.004i  \cr
                            -0.121+0.035i& -0.448-0.004i&0.683 }$,
                         $\pmatrix{0.157  &0.273-0.004i&0.233-0.055i \cr
                             0.273+0.004i & 0.476&0.408-0.090i         \cr
                             0.233+0.055i &0.408+0.090i&0.367 }$) \\ \hline
$\Gamma_6$  &   ($0.327542, 0.285361, 0.387097 \;;\;$
                          $\pmatrix{0.535  &-0.335+0.328i&0.020-0.168i \cr
                          -0.335-0.328i& 0.411&-0.116+0.093i\cr
                          0.020+0.168i &-0.116-0.093i&0.054 }$, \\
  &                    $\pmatrix{0.392& 0.375-0.250i& 0.152+0.113i \cr
                              0.375+0.250i&0.516 &0.073+0.204i \cr
0.152-0.113i  &0.073-0.204i&0.091 \cr},$
                          $\pmatrix{0.039 &-0.009-0.038i&-0.189+0.002i\cr
                           -0.009+0.038i&0.039&0.041-0.186i \cr
                           -0.189-0.002i &0.041+0.186i&0.922 })$\\ \hline
\end{tabular}
\end{center}\label{syujyotai}
\end{table*}%

We have observed that the additivity exactly holds for all the cases we examined, 
as is seen in Table \ref{results} for the representative examples. 
Table \ref{piast} shows the probability distributions which maximize the
quantum mutual information of the quantum channels $ \Gamma_i \; (i=1,
\cdots ,6)$. In the case of product channels,
the probability distribution $\displaystyle \pi^{\ast}
:=\argmax _{\pi}\I(\pi ; \Gamma ^{(1)} \otimes \Gamma ^{(2)})$ has turned
out to be
 the product probability distribution of
$\displaystyle
\pi_{1}^{\ast} = (\lambda_{i1}^{\ast} ; \sigma_{i1}^{\ast} )
:=\argmax _{\pi}  \I(\pi ; \Gamma ^{(1)} )$ and $\displaystyle
\pi_{2}^{\ast}=(\lambda_{j2}^{\ast} ; \sigma_{j2}^{\ast})
:=\argmax _{\pi}\I(\pi ; \Gamma ^{(2)} )$, which assigns probability
$\lambda_{i1}^{\ast}\lambda_{j2}^{\ast}$ to the state $\sigma_{i1}^{\ast}
\otimes
\sigma_{j2}^{\ast}$.

Fig.\ 1 illustrates 
the change in the quantum mutual information $\I(\pi^{(k)},
\Gamma^{(1)} \otimes \Gamma^{(2)})$ for $\Gamma^{(1)}=\Gamma^{(2)} =
\Gamma_2$
in the process of recursive computation $\pi^{(k)} \rightarrow \pi^{(k+1)}$
starting from some entangled states in ${\cal S}({\cal H}_1^{(1)}
\otimes
{\cal H}_1^{(2)})$. In addition, we measure the entanglement
\footnote{According to the criterion of \cite{Vedral}, the relative entropy
$\D(\sigma_{i}^{(k)} \|\sigma_{i1}^{(k)}\otimes  \sigma_{i2}^{(k)})$ is
inappropriate as a measure of entanglement in $\sigma_{i}^{(k)}$
since it takes a positive value even when
$\sigma_{i}^{(k)}$ is a classical mixture (convex combination)
of several product states.
Nevertheless, it does not mean that its use is inappropriate for our study.}
 of the states in
$\pi^{(k)}=(\lambda_{i}^{(k)};\sigma_{i}^{(k)})$ by
\begin{equation}
  \Ent(\pi^{(k)}):= \sum_{i}\lambda_{i}^{(k)}\D(\sigma_{i}^{(k)}
\|\sigma_{i1}^{(k)}\otimes  \sigma_{i2}^{(k)}),
\end{equation}
where $\sigma_{i1}^{(k)}$ and $ \sigma_{i2}^{(k)}$ are the marginal states
of
$\sigma_{i}^{(k)}$
defined by  partial trace.
Fig.\ 2 shows how the states get disentangled through the recursion.

\begin{figure}[t]
\begin{center}
\setlength{\unitlength}{0.120450pt}
\ifx\plotpoint\undefined\newsavebox{\plotpoint}\fi
\sbox{\plotpoint}{\rule[-0.200pt]{0.400pt}{0.400pt}}%
\begin{picture}(1500,900)(0,0)
\font\gnuplot=cmr10 at 10pt
\gnuplot
\sbox{\plotpoint}{\rule[-0.200pt]{0.400pt}{0.400pt}}%
\put(181,123){\makebox(0,0)[r]{0.5}}
\put(201.0,307.0){\rule[-0.200pt]{4.818pt}{0.400pt}}
\put(181,307){\makebox(0,0)[r]{0.505}}
\put(201.0,492.0){\rule[-0.200pt]{4.818pt}{0.400pt}}
\put(181,492){\makebox(0,0)[r]{0.51}}
\put(201.0,676.0){\rule[-0.200pt]{4.818pt}{0.400pt}}
\put(181,676){\makebox(0,0)[r]{0.515}}
\put(181,860){\makebox(0,0)[r]{0.52}}
\put(328.0,123.0){\rule[-0.200pt]{0.400pt}{4.818pt}}
\put(328,82){\makebox(0,0){5}}
\put(487.0,123.0){\rule[-0.200pt]{0.400pt}{4.818pt}}
\put(487,82){\makebox(0,0){10}}
\put(645.0,123.0){\rule[-0.200pt]{0.400pt}{4.818pt}}
\put(645,82){\makebox(0,0){15}}
\put(804.0,123.0){\rule[-0.200pt]{0.400pt}{4.818pt}}
\put(804,82){\makebox(0,0){20}}
\put(963.0,123.0){\rule[-0.200pt]{0.400pt}{4.818pt}}
\put(963,82){\makebox(0,0){25}}
\put(1122.0,123.0){\rule[-0.200pt]{0.400pt}{4.818pt}}
\put(1122,82){\makebox(0,0){30}}
\put(1280.0,123.0){\rule[-0.200pt]{0.400pt}{4.818pt}}
\put(1280,82){\makebox(0,0){35}}
\put(1439,82){\makebox(0,0){40}}
\put(201.0,123.0){\rule[-0.200pt]{149.117pt}{0.200pt}}
\put(1439.0,123.0){\rule[-0.200pt]{0.200pt}{88.772pt}}
\put(201.0,860.0){\rule[-0.200pt]{149.117pt}{0.200pt}}
\put(90,940){\makebox(0,0){$\I(\pi^{(k)}; \Gamma^{(1)}\otimes
\Gamma^{(2)})$}}
\put(820,21){\makebox(0,0){$k$}}
\put(201.0,123.0){\rule[-0.200pt]{0.200pt}{88.772pt}}
\put(233,253){\raisebox{-.8pt}{\makebox(0,0){$\Diamond$}}}
\put(264,591){\raisebox{-.8pt}{\makebox(0,0){$\Diamond$}}}
\put(296,666){\raisebox{-.8pt}{\makebox(0,0){$\Diamond$}}}
\put(328,705){\raisebox{-.8pt}{\makebox(0,0){$\Diamond$}}}
\put(360,728){\raisebox{-.8pt}{\makebox(0,0){$\Diamond$}}}
\put(391,741){\raisebox{-.8pt}{\makebox(0,0){$\Diamond$}}}
\put(423,750){\raisebox{-.8pt}{\makebox(0,0){$\Diamond$}}}
\put(455,755){\raisebox{-.8pt}{\makebox(0,0){$\Diamond$}}}
\put(487,758){\raisebox{-.8pt}{\makebox(0,0){$\Diamond$}}}
\put(518,760){\raisebox{-.8pt}{\makebox(0,0){$\Diamond$}}}
\put(550,761){\raisebox{-.8pt}{\makebox(0,0){$\Diamond$}}}
\put(582,762){\raisebox{-.8pt}{\makebox(0,0){$\Diamond$}}}
\put(614,762){\raisebox{-.8pt}{\makebox(0,0){$\Diamond$}}}
\put(645,762){\raisebox{-.8pt}{\makebox(0,0){$\Diamond$}}}
\put(677,762){\raisebox{-.8pt}{\makebox(0,0){$\Diamond$}}}
\put(709,762){\raisebox{-.8pt}{\makebox(0,0){$\Diamond$}}}
\put(741,763){\raisebox{-.8pt}{\makebox(0,0){$\Diamond$}}}
\put(772,763){\raisebox{-.8pt}{\makebox(0,0){$\Diamond$}}}
\put(804,763){\raisebox{-.8pt}{\makebox(0,0){$\Diamond$}}}
\put(836,763){\raisebox{-.8pt}{\makebox(0,0){$\Diamond$}}}
\put(868,763){\raisebox{-.8pt}{\makebox(0,0){$\Diamond$}}}
\put(899,763){\raisebox{-.8pt}{\makebox(0,0){$\Diamond$}}}
\put(931,763){\raisebox{-.8pt}{\makebox(0,0){$\Diamond$}}}
\put(963,763){\raisebox{-.8pt}{\makebox(0,0){$\Diamond$}}}
\put(995,763){\raisebox{-.8pt}{\makebox(0,0){$\Diamond$}}}
\put(1026,763){\raisebox{-.8pt}{\makebox(0,0){$\Diamond$}}}
\put(1058,763){\raisebox{-.8pt}{\makebox(0,0){$\Diamond$}}}
\put(1090,763){\raisebox{-.8pt}{\makebox(0,0){$\Diamond$}}}
\put(1122,763){\raisebox{-.8pt}{\makebox(0,0){$\Diamond$}}}
\put(1153,763){\raisebox{-.8pt}{\makebox(0,0){$\Diamond$}}}
\put(1185,763){\raisebox{-.8pt}{\makebox(0,0){$\Diamond$}}}
\put(1217,763){\raisebox{-.8pt}{\makebox(0,0){$\Diamond$}}}
\put(1249,763){\raisebox{-.8pt}{\makebox(0,0){$\Diamond$}}}
\put(1280,763){\raisebox{-.8pt}{\makebox(0,0){$\Diamond$}}}
\put(1312,763){\raisebox{-.8pt}{\makebox(0,0){$\Diamond$}}}
\put(1344,763){\raisebox{-.8pt}{\makebox(0,0){$\Diamond$}}}
\put(1376,763){\raisebox{-.8pt}{\makebox(0,0){$\Diamond$}}}
\put(1407,763){\raisebox{-.8pt}{\makebox(0,0){$\Diamond$}}}

\end{picture}
\caption[fig1]{The quantum mutual information $\I(\pi^{(k)};  \Gamma^{(1)}
\otimes
\Gamma^{(2)})$
in the process of the recursion $\pi^{(k)}\rightarrow\pi^{(k+1)}$}
\end{center}
\end{figure}
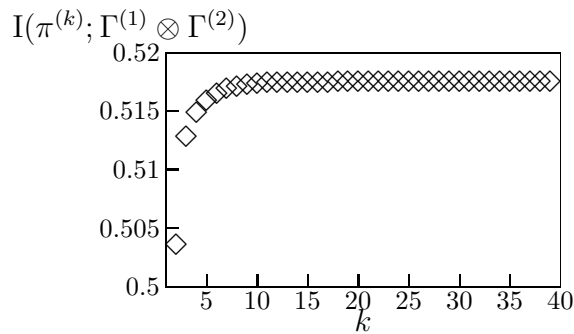
\begin{figure}
\begin{center}
\setlength{\unitlength}{0.120450pt}
\ifx\plotpoint\undefined\newsavebox{\plotpoint}\fi
\sbox{\plotpoint}{\rule[-0.200pt]{0.400pt}{0.400pt}}%
\begin{picture}(1500,900)(0,0)
\font\gnuplot=cmr10 at 10pt
\gnuplot
\sbox{\plotpoint}{\rule[-0.200pt]{0.400pt}{0.400pt}}%
\put(141,123){\makebox(0,0)[r]{-30}}
\put(161.0,246.0){\rule[-0.200pt]{4.818pt}{0.400pt}}
\put(141,246){\makebox(0,0)[r]{-25}}
\put(161.0,369.0){\rule[-0.200pt]{4.818pt}{0.400pt}}
\put(141,369){\makebox(0,0)[r]{-20}}
\put(161.0,492.0){\rule[-0.200pt]{4.818pt}{0.400pt}}
\put(141,492){\makebox(0,0)[r]{-15}}
\put(161.0,614.0){\rule[-0.200pt]{4.818pt}{0.400pt}}
\put(141,614){\makebox(0,0)[r]{-10}}
\put(161.0,737.0){\rule[-0.200pt]{4.818pt}{0.400pt}}
\put(141,737){\makebox(0,0)[r]{-5}}
\put(141,860){\makebox(0,0)[r]{0}}
\put(259.0,123.0){\rule[-0.200pt]{0.400pt}{4.818pt}}
\put(259,82){\makebox(0,0){2}}
\put(456.0,123.0){\rule[-0.200pt]{0.400pt}{4.818pt}}
\put(456,82){\makebox(0,0){4}}
\put(653.0,123.0){\rule[-0.200pt]{0.400pt}{4.818pt}}
\put(653,82){\makebox(0,0){6}}
\put(849.0,123.0){\rule[-0.200pt]{0.400pt}{4.818pt}}
\put(849,82){\makebox(0,0){8}}
\put(1046.0,123.0){\rule[-0.200pt]{0.400pt}{4.818pt}}
\put(1046,82){\makebox(0,0){10}}
\put(1242.0,123.0){\rule[-0.200pt]{0.400pt}{4.818pt}}
\put(1242,82){\makebox(0,0){12}}
\put(1439,82){\makebox(0,0){14}}
\put(161.0,123.0){\rule[-0.200pt]{153.935pt}{0.200pt}}
\put(1439.0,123.0){\rule[-0.200pt]{0.200pt}{88.772pt}}
\put(161.0,860.0){\rule[-0.200pt]{153.935pt}{0.200pt}}
\put(90,940){\makebox(0,0){$\log\Ent(\pi^{(k)})$}}
\put(800,21){\makebox(0,0){$k$}}
\put(161.0,123.0){\rule[-0.200pt]{0.200pt}{88.772pt}}
\put(161,750){\raisebox{-.8pt}{\makebox(0,0){$\Diamond$}}}
\put(259,567){\raisebox{-.8pt}{\makebox(0,0){$\Diamond$}}}
\put(358,409){\raisebox{-.8pt}{\makebox(0,0){$\Diamond$}}}
\put(456,323){\raisebox{-.8pt}{\makebox(0,0){$\Diamond$}}}
\put(554,304){\raisebox{-.8pt}{\makebox(0,0){$\Diamond$}}}
\put(653,291){\raisebox{-.8pt}{\makebox(0,0){$\Diamond$}}}
\put(751,279){\raisebox{-.8pt}{\makebox(0,0){$\Diamond$}}}
\put(849,267){\raisebox{-.8pt}{\makebox(0,0){$\Diamond$}}}
\put(947,255){\raisebox{-.8pt}{\makebox(0,0){$\Diamond$}}}
\put(1046,243){\raisebox{-.8pt}{\makebox(0,0){$\Diamond$}}}
\put(1144,231){\raisebox{-.8pt}{\makebox(0,0){$\Diamond$}}}
\put(1242,219){\raisebox{-.8pt}{\makebox(0,0){$\Diamond$}}}
\put(1341,207){\raisebox{-.8pt}{\makebox(0,0){$\Diamond$}}}
\put(1439,183){\raisebox{-.8pt}{\makebox(0,0){$\Diamond$}}}

\end{picture}
\caption[fig2]{Semi-logarithmic plot of entanglement versus iteration
number}
\end{center}
\end{figure}
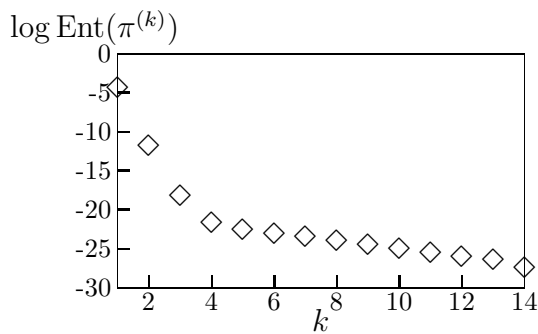

\section{Conclusions}
We have applied the quantum Arimoto-Blahut algorithm 
to various quantum channels $\Gamma^{(1)}$ and 
$\Gamma^{(2)}$ 
with 
${\cal H}_1 ={\cal H}_2 ={\bf C}^2$ and 
${\cal H}_1 ={\cal H}_2 ={\bf C}^3$ 
as well as their product channels 
$\Gamma^{(1)}\otimes\Gamma^{(2)}$, 
and have verified that the additivity (\ref{additive}) 
holds for all the examples investigated.  
Note that the additivity (\ref{additive}) has 
been proved only for some special classes of 
channels so far as explained in section 
\ref{sec:related_works} and that 
the examined channels do not 
belong to them.  
Needless to say, this is not a theoretical analysis but 
a numerical one, applied to only a limited number of channels 
on low-dimensional Hilbert spaces, using   
an algorithm which does not ensure the  global maximum.  Therefore we cannot 
rely upon the obtained results too much. 
Nevertheless, it seems very 
unlikely that all the randomly chosen examples happen to 
satisfy the additivity by a coincidence or that 
dimensions $2$ and $3$ are special in a property like 
the additivity which is not explicitly related to 
the dimension.  We are thus naturally led to 
conclude that the results  suggest 
that the additivity always holds. 

\section*{Acknowledgement}
We would like to thank Dr.\ A.S.\ Holevo of Steklov Mathematical Institute
and
Dr.\ A.\ Fujiwara of the Osaka
University for giving crucial comments on this work.

\appendix
\section{Proof of Theorem 1}
  Since $C(\Gamma ^{\otimes (N+M)}) \ge C(\Gamma ^{\otimes N})+C(\Gamma
^{\otimes
M})$ holds, $\displaystyle \lim_{N \to \infty}\frac{C(\Gamma ^{\otimes
N})}{N}$ exists and is proved to be $\displaystyle \sup_{N}\frac{C(\Gamma
^{\otimes
N})}{N}$.
Let $\{\Phi_N\}_{N=1}^{\infty}$ be a sequence of coding systems
satisfying $\displaystyle  \lim_{N \to \infty} P_{er}( \Phi_N ,
\Gamma ) = 0$,
 where $\Phi_N$ consists of
codewords $\{\sigma^{(N)}(i)\}_{i=1}^{M_N}$ which are arbitrary states on
${\cal
H}_1^{\otimes
N}$ and a measurement $X^{(N)} = \{X^{(N)}_i\}_{i=1}^{M_N}$.
Let $Y_N$ be the classical
random variable on the message set $\{1, \cdots, M_N\}$ which
takes each value with
equal probability $1/M_N$, and let $\hat {Y}_N$ be the classical random
variable on the same set which represents the decoded message obtained by
performing the measurement $X^{(N)}$ to the output state
$\Gamma^{\otimes N} (\sigma^{(N)}(i))$, where
the transmitted message $i$ is assumed to be $Y_N$.
Then the Fano inequality (see e.g. \cite{Cover}) implies
\[1+P_{er} ( \Phi_N , \Gamma )\log M_N \ge \log M_N-\I(Y_N ; \hat{Y}_N), \]
where $\I(Y_N ; \hat{Y}_N)$ is the classical mutual information
between $Y_N$ and $\hat{Y}_N$.
This leads to
\begin{equation}
(1-P_{er}( \Phi_N , \Gamma ))\frac{1}{N} \log M_N \le \frac{1}{N}+
\frac{1}{N}\I(Y_N ; \hat{Y}_N).
 \label{fano2} \end{equation}
In addition, we have
\begin{eqnarray}
\lefteqn{\I(Y_N;\hat{Y}_N)} \nonumber\\
&&= \frac{1}{M_N} \sum_{i=1}^{M_N}\D_{X^{(N)}}(\Gamma ^{\otimes
N}(\sigma^{(N)}(i))\| \Gamma
^{\otimes N}(\rho^{(N)} ))  \nonumber\\
&& \le \frac{1}{M_N} \sum_{i=1}^{M_N} \D (\Gamma ^{\otimes
N}(\sigma^{(N)}(i))\|\Gamma
^{\otimes N}(\rho^{(N)} ))\nonumber \\
&&\le \max_{\pi} \I(\pi;\Gamma ^{\otimes N}) \nonumber \\
  &&= C(\Gamma ^{\otimes N}), \label{cgamma}
\end{eqnarray}
where $\D_{X^{(N)}}(\Gamma ^{\otimes N}(\sigma^{(N)}(i))\| \Gamma ^{\otimes
N}
( \rho^{(N)} ))$ is the classical relative entropy between the conditional
probability
$P_{\hat{Y}_N|Y_N} (\cdot|i ):=\tr [\Gamma ^{\otimes
N}(\sigma^{(N)}(i))X.^{(N)}]$
and the probability $P_{\hat{Y}}(\cdot):=\tr[\Gamma ^{\otimes
N}(\rho^{(N)})X.^{(N)}]$ with
$\rho^{(N)}:=\frac{1}{M_N}\sum_{i}\sigma^{(N)}(i)$, and
the first inequality follows from the monotonicity of the relative entropy.
Substituting (\ref{cgamma}) into (\ref{fano2}), letting N $\to \infty $ and
taking
supremum with respect to $\{ \Phi_N\}_{N=1}^{\infty}$, we come to the
inequality $\displaystyle
\tilde{C}(\Gamma) \le  \lim_{N \to \infty} \frac{C(\Gamma ^{\otimes
N})}{N}$.
Conversely, since $C(\Gamma ^{\otimes N})$ is the supremum of the limit
values of
the rates of asymptotically error-free coding systems whose codewords are
restricted
to product states  of the form $\rho_{1}\otimes  \cdots \otimes \rho_{n} \in
{\cal
S}({\cal H}_1^{\otimes Nn})$, where $\rho_{i}\;(i=1,\cdots ,n)$ are
arbitrary states
on ${\cal H}_{1} ^{\otimes N}$,
it cannot be greater than $N \tilde{C}(\Gamma )$ by the definition of
$\tilde{C}(\Gamma )$.
 Hence we have
$\displaystyle \tilde{C}(\Gamma) \ge  \lim_{N \to \infty} \frac{C(\Gamma
^{\otimes
N})}{N}$.
\QED
\section{Operator-sum representation}
An arbitrary  completely positive trace preserving linear map
$\Gamma \, : \,{\cal T}({\cal H}_{1})\to{\cal
T}({\cal H}_{2})$ can be written in the form
\[ \Gamma (\rho)=\sum_{k=1}^m V_k \rho V_k^{\ast} \]
where ${\cal V}=\{V_k\}_{k=1}^m$ is a collection of bounded operators
from ${\cal H}_1$ to ${\cal H}_2$
satisfying
$\sum_{k=1}^m V_k^{\ast}V_k =\I$ \cite{Kraus} and $m$ can be taken at most $\dim {\cal H}_1\dim {\cal H}_2$ \cite{fujiwara2}. This is called the {\it
operator-sum
representation} or the {\it Kraus decomposition} of $\Gamma$
with the {\it generator} ${\cal V}$.

\section{Quantum binary channel}
A quantum channel whose input and output systems are both
${\bf C}^2 $
is called a {\it quantum binary channel}.  Since every
density operator on ${\bf C}^2 $ is
represented in the form
\[ \rho_{\theta} = \frac{1}{2} \pmatrix{
            1+\theta_3           &  \theta_1 - i\theta_2 \cr
            \theta_1+  i\theta_2  &    1-\theta_3        \cr
            } \]
with $\theta = (\theta_1, \theta_2, \theta_3)^t $
lying in  the unit ball
\[   {\cal V}=\{ \theta  \in {\bf R}^3\; ;\; \parallel\theta
\parallel^2 = \theta_1^2 + \theta_2^2 +\theta_3^2 \le 1\},  \]
an arbitrary quantum binary channel is represented as
$\Gamma(\rho_{\theta})=\rho_{A
\theta +b}$ by a  $3
\times 3 $ real matrix $A$ and a 3-dimensional real column vector $b$.
We denote such a channel by $\Gamma = (A,b)$.
For representing a completely positive map,
they should satisfy the following condition \cite{fujiwara2}

\[ \pmatrix{ 
        \frac{1}{2}+p  &  x            &  r            & w  \cr
        \overline{x}   & \frac{1}{2}-p & y             & -r  \cr
        \overline{r}   & \overline{y}  & \frac{1}{2}+q & z    \cr
        \overline{w}   & -\overline{r} & \overline{z} & \frac{1}{2}-q \cr
                                                               }  \ge 0  \]
when A and b are represented as
    \[  A=  \pmatrix{
                y_R +w_R  &   y_I+ w_I   &  x_R-z_R \cr
                 y_I - w_I &   -y_R+w_R  &  -x_I+ z_I \cr
                 2r_R      &   2r_I      &   p-q      \cr
                                                   },  \]
   \[    b= \pmatrix{
                   x_R+z_R \cr
                  -x_I-z_I \cr
                   p+q      \cr
                     }.   \]   
(The subscripts $R$ and $I$ denote the real and imaginary parts, i.e. $x=x_R+i x_I$, 
etc.)

\end{document}